\documentclass[12pt,preprint]{aastex}

\newcommand{\CXOJ}{CXO~J192318.5$+$143035}

\slugcomment{Accepted for publication in ApJL}

\shorttitle{Gamma-ray Study of SNR W51C}
\shortauthors{Fermi LAT Collaboration}

\begin{document}


\title{Fermi-LAT Discovery of Extended Gamma-ray Emission \\
in the Direction of Supernova Remnant W51C}

\author{
A.~A.~Abdo\altaffilmark{2,3}, 
M.~Ackermann\altaffilmark{4}, 
M.~Ajello\altaffilmark{4}, 
L.~Baldini\altaffilmark{5}, 
J.~Ballet\altaffilmark{6}, 
G.~Barbiellini\altaffilmark{7,8}, 
M.~G.~Baring\altaffilmark{9}, 
D.~Bastieri\altaffilmark{10,11}, 
B.~M.~Baughman\altaffilmark{12}, 
K.~Bechtol\altaffilmark{4}
% ...
}
% ...
\altaffiltext{2}{Space Science Division, Naval Research Laboratory, Washington, DC 20375, USA}
\altaffiltext{3}{National Research Council Research Associate, National Academy of Sciences, Washington, DC 20001, USA}
\altaffiltext{4}{W. W. Hansen Experimental Physics Laboratory, Kavli Institute for Particle Astrophysics and Cosmology, Department of Physics and SLAC National Accelerator Laboratory, Stanford University, Stanford, CA 94305, USA}
\altaffiltext{5}{Istituto Nazionale di Fisica Nucleare, Sezione di Pisa, I-56127 Pisa, Italy}
\altaffiltext{6}{Laboratoire AIM, CEA-IRFU/CNRS/Universit\'e Paris Diderot, Service d'Astrophysique, CEA Saclay, 91191 Gif sur Yvette, France}
\altaffiltext{7}{Istituto Nazionale di Fisica Nucleare, Sezione di Trieste, I-34127 Trieste, Italy}
\altaffiltext{8}{Dipartimento di Fisica, Universit\`a di Trieste, I-34127 Trieste, Italy}
\altaffiltext{9}{Rice University, Department of Physics and Astronomy, MS-108, P. O. Box 1892, Houston, TX 77251, USA}
\altaffiltext{10}{Istituto Nazionale di Fisica Nucleare, Sezione di Padova, I-35131 Padova, Italy}
\altaffiltext{11}{Dipartimento di Fisica ``G. Galilei", Universit\`a di Padova, I-35131 Padova, Italy}
\altaffiltext{12}{Department of Physics, Center for Cosmology and Astro-Particle Physics, The Ohio State University, Columbus, OH 43210, USA}
% ...

\begin{abstract}
The discovery of bright gamma-ray emission coincident with supernova remnant (SNR) W51C is reported 
using the Large Area Telescope (LAT) 
on board the Fermi Gamma-ray Space Telescope. 
W51C is a middle-aged remnant ($\sim 10^4$ yr) 
with intense radio synchrotron emission in its shell and 
known to be interacting with a molecular cloud. 
The gamma-ray emission is spatially extended, broadly consistent with 
the radio and X-ray extent of SNR W51C. The energy spectrum 
in the 0.2--50 GeV band exhibits steepening toward high energies.
The luminosity 
is greater than $1\times 10^{36}\ \rm erg\ s^{-1}$ 
given the distance constraint of $D>5.5$ kpc, which 
makes this object one of the most luminous gamma-ray sources 
in our Galaxy. 
The observed  gamma-rays  can be explained reasonably by a combination of  
efficient acceleration of nuclear cosmic rays at supernova shocks and 
shock-cloud interactions. 
The decay of neutral $\pi$-mesons produced in hadronic collisions
provides a plausible explanation for the gamma-ray emission. 
The product of the average gas density and the total energy content of 
the accelerated protons amounts to  
$\bar{n}_{\rm H}W_p \simeq 5\times 10^{51}\ (D/6\ {\rm kpc})^2\ 
\rm erg\ cm^{-3}$.
Electron density constraints from the radio and X-ray bands render it difficult to explain the LAT signal as due to inverse Compton scattering. 
The \emph{Fermi} LAT source coincident with 
SNR W51C sheds new light on the origin of Galactic cosmic rays. 
\end{abstract}


\section{Introduction}

The origin of cosmic rays  remains unsolved.
The idea that supernovae produce cosmic rays  \citep[e.g.,][]{Hayakawa56}
has been developed both in theoretical and observational aspects, 
so that galactic cosmic rays 
are widely considered to be accelerated in the expanding shocks
of supernova remnants (SNRs) through diffusive shock acceleration 
process \citep[see e.g.,][]{BE87}.
This conjecture has been strengthened by
recent observations of young SNRs in synchrotron X-rays and very-high-energy (VHE) 
gamma-rays \citep[see e.g.,][]{Reynolds08,Aha08}.
High efficiency for converting the kinetic energy released by supernova explosions 
 into the energy of 
relativistic protons and nuclei is a key requirement to explain the galactic cosmic rays; 
it is yet to be confirmed observationally. 

Enhanced $\pi^0$-decay emission expected 
from SNRs that are interacting with  molecular clouds 
could provide direct evidence of 
 the nuclear component of cosmic rays being accelerated at supernova shocks \citep{ADV94}. 
To identify 
the  $\pi^0$-decay gamma-rays as evidence of the nuclear cosmic rays, 
observations in the GeV domain are crucial because the  $\pi^0$-decay spectrum 
has a characteristic  bump around 70 MeV.
Previous  measurements of GeV gamma-rays with EGRET onboard the Compton 
Gamma-Ray Observatory found some gamma-ray sources near 
radio-bright SNRs \citep{Esp96}. However, 
 the possible origins of the EGRET sources, namely SNR shell contributions 
or pulsar associations, remained unclear, mainly due to poor localization.

The advent of the Large Area Telescope (LAT) onboard 
the \emph{Fermi} Gamma-ray Space Telescope 
provides a new opportunity to study the gamma-ray emission 
from SNRs at GeV energies. 
An initial source list 
based on the first three months of \emph{Fermi} observations \citep{BSL} 
includes 0FGL~J1923.0$+$1411, which 
is spatially coincident with SNR W51C. 
Even with the three months data, the detection was at $\sim 23\sigma$ level. 
There are no EGRET counterpart(s) to the LAT source \citep{EGRET3}; 
this is the first SNR 
discovered by \emph{Fermi} as confirmed in this paper. 

W51C (G49.2$-$0.7) is a radio-bright SNR at a distance of 
$D \simeq 6$ kpc with an estimated age of $\sim 3\times 10^4$ yrs \citep{Koo95}.
It has an elliptical shape 
with an extent of $50\arcmin \times 38\arcmin$.
The radio continuum map exhibits  thick shell-like structures \citep{330MHz}.
The bulk of the X-ray emission comes from 
thermal plasma with a temperature of $kT\sim 0.3\ {\rm keV}$ \citep{Koo05}.
A molecular cloud--shock interaction is known, as evidenced by 
observations of shocked atomic and molecular gases \citep{KM97a,KM97b}.
Most recently, the HESS collaboration has announced the detection 
of  extended VHE gamma-ray emission coincident with W51C  \citep{HESSW51C}. 
Also, the Milagro collaboration 
has reported a possible excess of multi-TeV 
gamma-rays in this direction \citep{Milagro09}.

In this paper we report the analysis results for the LAT 
source coincident with SNR W51C,
using data accumulated over the first year of 
\emph{Fermi}'s operation. 
The \emph{Fermi} observations of SNR W51C  permit a refined study of cosmic-ray acceleration. Specifically, the LAT data suggest that $\pi^0$-decay emission is the
dominant contribution to the gamma-ray signal.

\section{Observation and Data Reduction}

The \emph{Fermi} Gamma-ray Space Telescope was launched 
on 2008 June 11 by a Delta II Heavy launch vehicle. 
The LAT onboard \emph{Fermi}
 is a pair-conversion gamma-ray detector capable of measuring 
gamma-rays in a very wide range of energy from 20 MeV up to 300 GeV. 
The LAT tracks the electron and positron  resulting from pair conversion of 
an incident gamma-ray in thin high-$Z$ foils, and measures
the energy deposition due to the subsequent electromagnetic shower that develops in the calorimeter.
The effective area is $\sim 8000\ {\rm cm^{2}}$ above 1 GeV
(on-axis) and the per-photon 68\% containment radius is  $\sim 0\fdg 8$ at 1 GeV. 
The point-spread function (PSF) depends largely on  photon energy 
and improves  at higher energies. 
The tracker of the LAT is divided into two regions, \emph{front} and \emph{back}.
The front region (first 12 planes) has thin converters 
to optimize the PSF while the back region (4 planes after 
the front section) has thicker converters to enlarge the effective area.
The angular resolution for the 
back events is approximately twice as broad than that for the front events.
The details of the LAT  and data processing are given in \citet{LAT}, 
and the on-orbit calibration is described in \citet{Onorbit}.

The gamma-ray data acquired from 2008 August 5 to  2009 July 14 are 
analyzed. 
The \emph{diffuse} class events as defined in \citet{LAT} 
are chosen for gamma-ray analysis. 
A cut on earth zenith angles greater than $105^{\circ}$ 
is applied to reduce the residual signal from earth albedo gamma-rays. 
 The instrument response functions (IRFs) called ``Pass 6 V3" 
are used \citep{Rando}.

\section{Analysis and Results}

The maximum likelihood technique is employed for spectral 
and spatial parameter estimation using 
{\it gtlike}, which is publicly available as 
part of  \emph{Fermi} Science 
Tools\footnote{Software and documentation of the \emph{Fermi} Science Tools are distributed by the Fermi Science Support Center at http://fermi.gsfc.nasa.gov/ssc}. 
The likelihood is the product of the probability of observing the gamma-ray counts  
of each spatial and energy bin given the emission model,
and
parameter values are estimated by maximizing the likelihood of the data 
given the model \citep{Mattox96}.
The gamma-ray emission model includes individual sources at fixed coordinates, 
galactic diffuse emission (resulting 
from cosmic-ray interactions with interstellar gas and photons), 
and an isotropic  component (extragalactic and residual background). 
The so-called ``mapcube" file of  gll\_iem\_v02.fit 
is used for modeling the galactic diffuse emission, 
together with 
the corresponding tabulated model for the isotropic diffuse emission.
Other versions of the galactic diffuse models, 
generated by the GALPROP code \citep{GALPROP1},  
are also utilized to assess systematic error. 
The maximum likelihood analysis is performed inside a square region
of $12\degr \times 12\degr$ centered on W51C 
with a pixel size of $0\fdg 1$,  
unless otherwise mentioned. 
Background point sources detected in six months data 
are included in the likelihood analysis with free normalization and power-law index, 
though none of them affect  the results in this paper. 

\subsection{Spatial Distribution}

\begin{figure}[htbp] 
\epsscale{1.1}
\plottwo{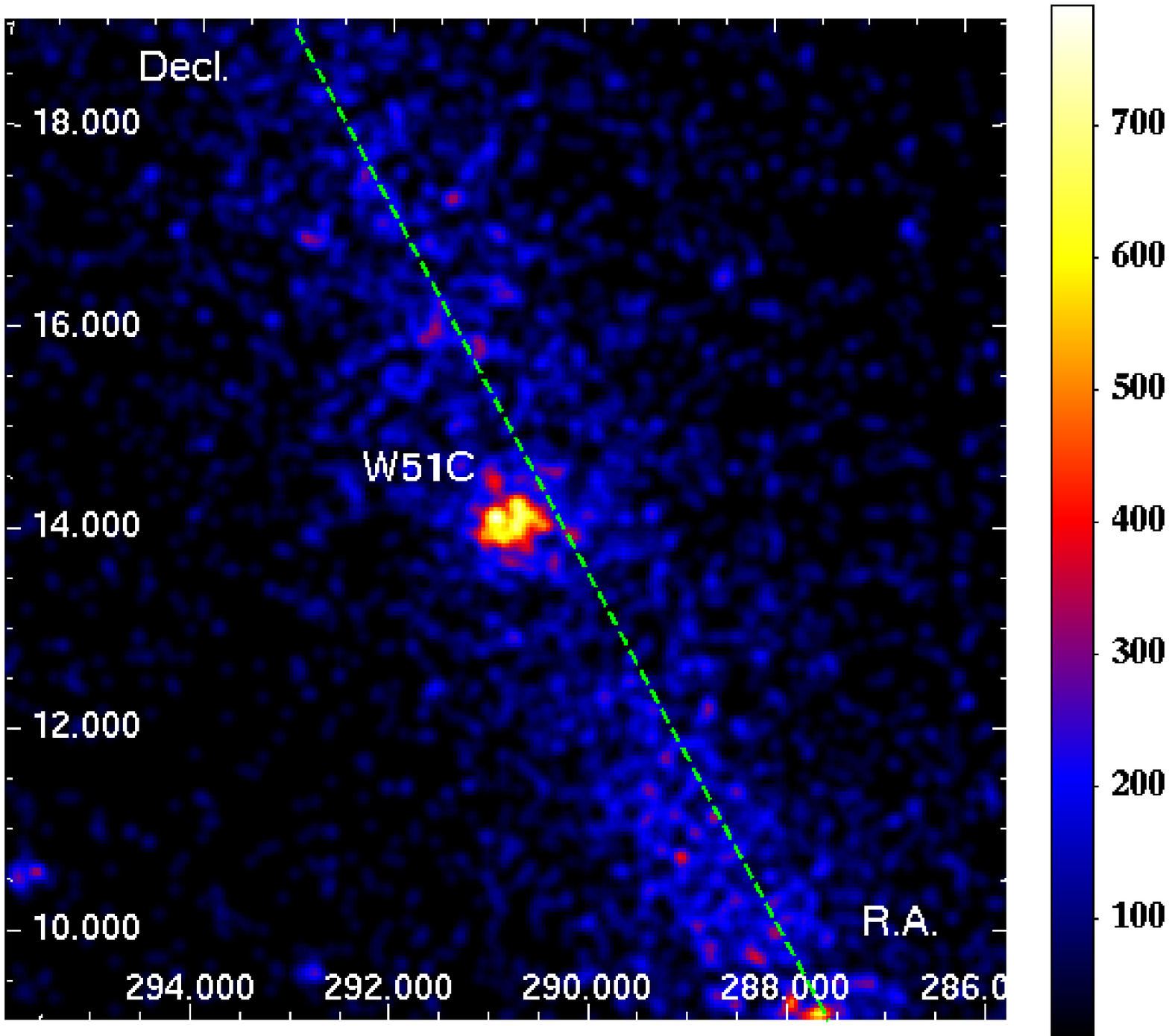}{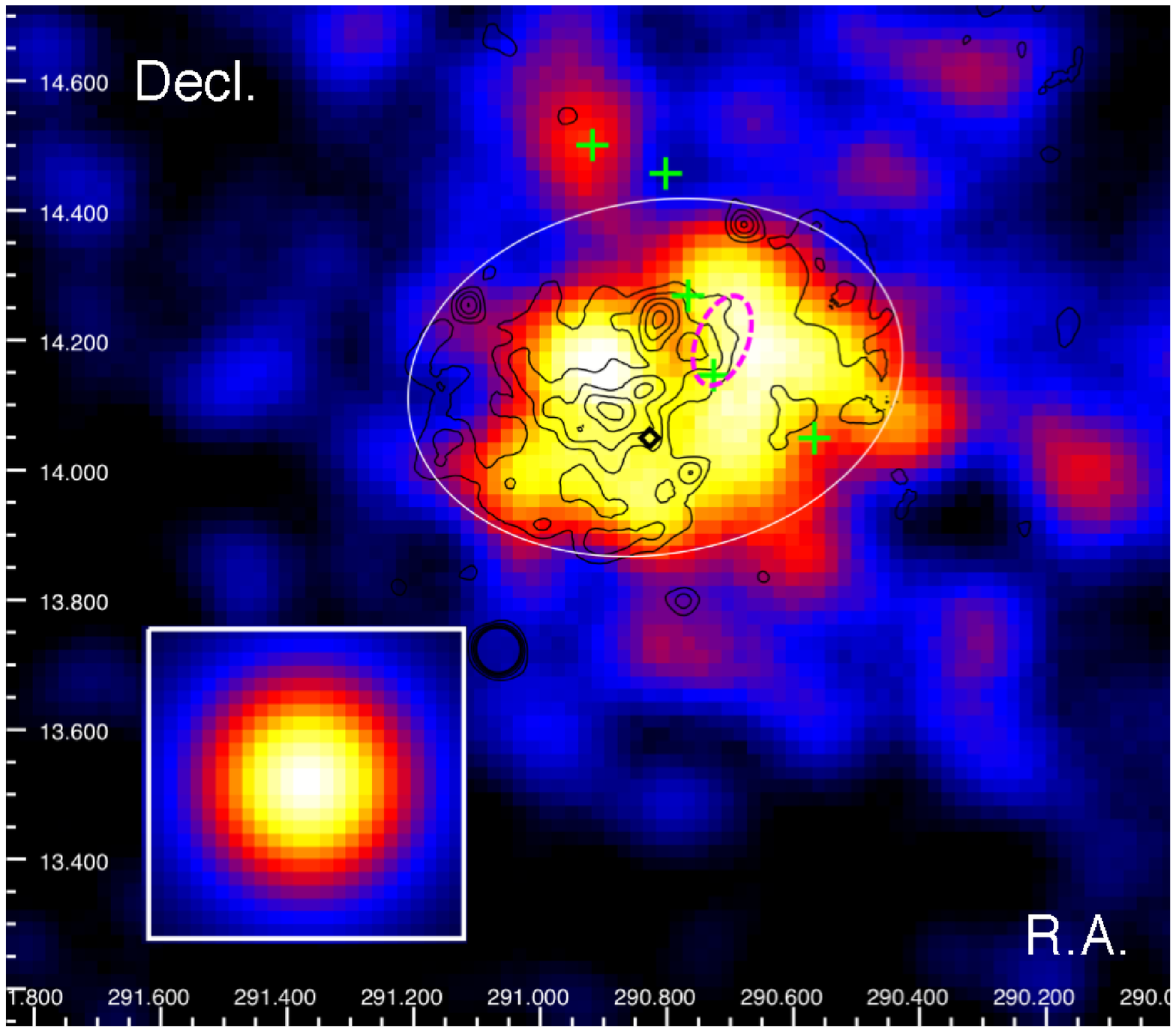}
\caption{(Left) \emph{Fermi} LAT counts map in 2--10 GeV around SNR W51C 
in units of counts per deg$^2$. Front-converted events are selected. 
The counts map is smoothed by a Gaussian kernel of $\sigma =0\fdg 12$.
The green dashed line represents the galactic plane.
(Right)
Close up view around SNR W51C.
The simulated point source image (smoothed by the same gaussian) 
is shown in the inset. 
The outer boundary of W51C is indicated by a white ellipse. 
Superposed is the ROSAT X-ray map (contours) from \citet{Koo95}. 
The region where shocked CO clumps \citep{KM97b} were found is represented by 
a dashed magenta ellipse. 
A diamond near the SNR centroid indicates  \CXOJ\ (see the text). 
The positions of five HII regions are indicated by green crosses \citep{CS98}. 
  \label{CountsMap}}
\end{figure}

In Figure~\ref{CountsMap}, the maps of photon counts 
in the 2--10 GeV band in the vicinity of  SNR W51C are shown; 
the right panel is a close-up view of the left panel. 
Gamma-ray events that 
converted in the \emph{front} section of the tracker are selected. 
A bright gamma-ray source is enclosed by the outer boundary of W51C.
The average surface brightness inside the SNR boundary 
is about 2 and 5 times larger than neighboring regions on the galactic plane
in the 0.5--2 GeV and 2--10 GeV bands, respectively.
The gamma-ray distribution appears to be somewhat clumpy, 
suggesting the presence of sub-structures. 
Due to the limited statistics, we defer the investigation of possible sub-structures 
to a future publication. 

\begin{figure}[htbp]
  \epsscale{0.7}
 \plotone{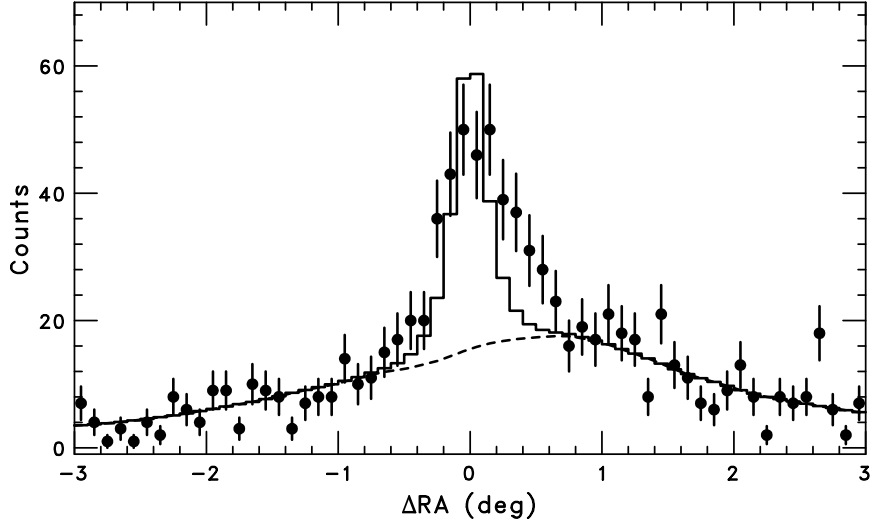}
  \caption{One dimensional profile of the observed 2--10 GeV gamma-rays 
(data points) along right ascension.
The counts map  is integrated 
over the direction of declination with a width of $\Delta  \delta = 1\fdg 6$ 
centered on W51C.
The dashed curve shows the profile of the galactic+isotropic diffuse model. 
The histogram represents the sum of a simulated point source and the diffuse model. 
  \label{fig:profile} }
 \end{figure}

The spatial distribution of the gamma-ray source 
is significantly extended compared to a simulated point source 
that has the same spectrum. 
A one dimensional profile in the right ascension direction 
of the counts map is compared with  that expected for a point source 
in Figure~\ref{fig:profile}.
Though the width of the PSF above 5 GeV is known to deviate from the Monte Carlo 
simulations,  it  has negligible effects on the simulated point source and 
other results in this paper. 
Assuming a two-dimensional gaussian distribution
fixed at the W51C centroid 
 ($\alpha$, $\delta$)=($290\fdg 818$, $14\fdg 145$), we 
calculate the extension parameter of the source as 
$\sigma_{\rm ext} = 0\fdg 22\pm 0\fdg 02$ by varying $\sigma_{\rm ext}$ to find 
the best match with the data. 
Note however that a different assumption on the spatial distribution results in 
different quantification of the spatial extension. 

The extended nature of this LAT source cannot be ascribed to 
a superposition of two point-like sources. To quantify this, we define a grid of 
60 positions inside the remnant as trial point source positions, 
 and calculate maximum likelihood values ($L_{\rm 2ps}$) 
for each possible combination 
by running {\it gtlike} for a box of $8\degr \times 8\degr$.
On the other hand, 
assuming that the gamma-ray source has uniform surface brightness 
inside the SNR boundary, 
we obtain another maximum likelihood value ($L_{\rm uni}$).
The resulting values of the likelihood test statistics, 
$-2\ln (L_{\rm 2ps}/L_{\rm uni}) > 32$, 
demonstrate that the uniform emission gives a much better result.

The extension of the source precludes magnetospheric radiation from a pulsar 
 as a dominant gamma-ray source. 
 However, a small fraction ($\sim 10\%$) of the gamma-ray 
flux could be contributed by a pulsar. 
Our pulse  search results in non-detection, 
which places a $5\sigma$ upper limit on the pulsed gamma-rays as 
$\simeq 6\times 10^{-8}\ {\rm photons\ cm^{-2}\ s^{-1}}$ above 100 MeV
\citep[see][]{PulsarCatalog}.

\subsection{Spectrum and Its Uncertainties}\label{sec:spectrum}

The gamma-ray spectrum of the W51C source  is shown in Figure~\ref{fig:SED}. 
It is obtained by performing likelihood analysis for each energy bin 
with {\it gtlike}. 
The lower energy bound is set at 0.2 GeV, below which 
the systematic uncertainties become too large due to the background diffuse emission 
(see below). 
The surface brightness of the W51C source is assumed to be uniform 
inside the SNR boundary. 
The normalization of the galactic diffuse emission model is left free 
at each energy bin. 
As evident from Figure~\ref{fig:SED}, 
the gamma-ray spectrum cannot be described by a simple power law and 
steepens above a few GeV. 
A likelihood-ratio test indicates that 
a power-law hypothesis has a chance probability of $5\times 10^{-5}$ 
for obtaining the spectral data.  The physical interpretation of the spectral 
energy distribution will be discussed in \S\ref{sec:discuss}.

The gamma-ray luminosity can be estimated 
as $\simeq 1\times 10^{36}\ (D/6\ {\rm kpc})^2\ {\rm erg\ s^{-1}}$ in the LAT bandpass, making this object one of the most luminous gamma-ray sources 
in our Galaxy. 
Note that the distance to the remnant is well constrained. 
X-ray absorption \citep{Koo95} indicates that 
W51C is situated behind the W51 molecular cloud complex 
(a so-called 68 ${\rm km\ s^{-1}}$ cloud in particular) 
whose distance is determined as $7.0\pm 1.5$ kpc by 
the proper motion of W51 Main $\rm H_2O$ maser \citep{Genzel81}. 
On the other hand, the angular extent and the X-ray measurements \citep{Koo02} 
require $D\la 10$~kpc. 
We adopt $D=6$~kpc following previous publications 
\citep[e.g.,][]{Koo05}.

It should be noted that 
in addition to the systematic errors commonly assigned to 
LAT spectral data\footnote{For the IRF used here
(P6\_V3\_DIFFUSE), systematic error as a function of  $x=\log (E/{\rm MeV})$
is 10\% at $x=2$, 5\% at $x=2.75$, 
and 20\% at $x=4$ with a linear interpolation between them. },
the accuracy of the flux estimated for 
 each energy bin of the W51C source 
is limited by the following errors. 
First, a possible effect of the uncertainty of 
underlying galactic diffuse emission is considered. 
The observed gamma-ray intensity of nearby source-free 
regions on the galactic plane is compared 
with the intensity expected from the galactic diffuse models. 
The difference, namely the local departure from the best-fit diffuse model, 
is found to be $\la 6\%$.
By changing 
the normalization of the galactic diffuse model artificially by $\pm 6\%$, 
we estimate the possible systematic error 
to be 40\% (0.2--0.4 GeV), 22\% (0.4--0.8 GeV) and $< 10\%$ ($>0.8$ GeV). 

The fact that we do not know the true gamma-ray morphology of the W51C 
source introduces another 
error in our flux estimation. Since the gamma-ray source is point-like 
below 1 GeV given the PSF, 
this uncertainty matters only above 1 GeV. 
We adopt a uniform surface brightness inside 
the SNR boundary (a flat elliptical template) 
 as the spatial distribution of the source gamma-rays.
Different spatial distributions such as a flat elliptical 
template reduced in size (scaled by 0.5) 
are tested to estimate the systematic error. 
Our conservative estimate is  $\la 20\%$  in 1--6 GeV 
and $\sim 30\%$ above 6 GeV as the systematic uncertainty attributable 
to the unknown shape of the source. 
 
\section{Discussion}\label{sec:discuss}

\begin{figure}[htbp]
  \epsscale{0.8}
    \plotone{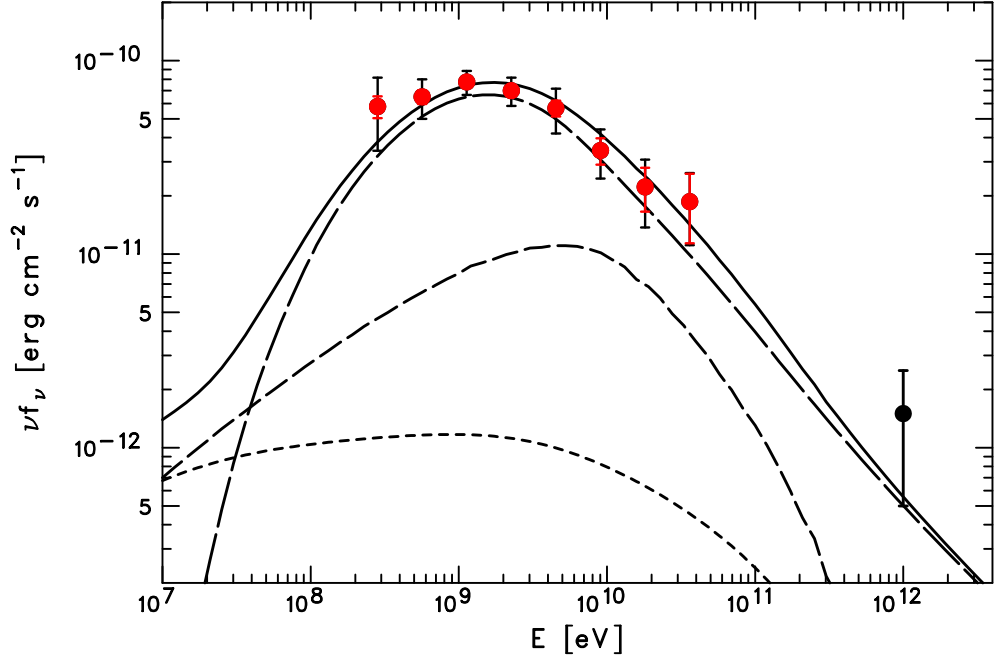}
  \caption{SED of the SNR W51C region measured with the \emph{Fermi} LAT 
(red points)  together with phenomenological modeling. 
Systematic errors (see \S\ref{sec:spectrum}) are indicated by black bars. 
  The model consists of $\pi^0$-decay (long-dashed curve), 
bremsstrahlung (dashed curve), and IC scattering (short-dashed curve).
The integrated flux reported by HESS is converted to 
the differential flux at 1 TeV assuming 
photon index $\Gamma =2.5\pm 1.0$ (black point). 
  \label{fig:SED}}
 \end{figure}

The extended gamma-ray emission positionally coincident with SNR W51C 
has been studied using the \emph{Fermi} LAT.
The gamma-ray spectrum presented in Figure~\ref{fig:SED} 
is not fit by a simple power law, exhibiting a remarkable steepening. 
 Here we discuss the origin of the extended emission and 
 the underlying 
particle spectra that give rise to the observed  spectrum of photons.

The expanding shock waves driven by the supernova explosion 
are expected to be the sites of the acceleration of multi-GeV particles.
To phenomenologically interpret the spectral curvature in the LAT+TeV bands,
a broken power-law is adopted for the momentum distribution of the radiating 
electrons/protons:
\begin{equation}
\label{eq:N}
   N_{e,p}(p) = a_{e,p} \left(\frac{p}{p_0}\right)^{-s} 
   \left(1+  \left(\frac{p}{p_{\rm br}}\right)^{2} \right)^{-\Delta s/2},
\end{equation}
where $p_0 = 1\ {\rm GeV}\, c^{-1}$.
For simplicity, 
the indices and the break momentum are assumed to be
 identical for both accelerated protons and electrons. 
As we argue below, the break may reflect the character 
of magnetohydrodynamic turbulence.
To account for  the radio synchrotron index $\alpha \simeq 0.26$ \citep{MK94}, 
we adopt $s=1.5$, though a steeper index (say, $s = 1.7$) 
could be reconciled  with the radio observations within the uncertainty. 
The energetic particles are assumed to be uniformly distributed in 
the volume of $V=(4\pi /3) R_{\rm eff}^3$ with an effective radius $R_{\rm eff}=30$ pc. 
The age and radius imply a shock velocity of 
$v_{\rm sh} \sim 400\ {\rm km\ s^{-1}}$ and 
$E_{\rm SN}/n_0 \sim 1.6\times 10^{52}\ {\rm erg\ cm^{3}}$, 
where $E_{\rm SN}$ and $n_0$ represent the explosion kinetic energy and 
the interstellar density into which the main blast wave is propagating,
respectively. The X-ray data suggest $n_0 \sim 0.3\ {\rm cm^{-3}}$.
The radio images indicate that 
radio-emitting electrons are smoothly distributed in a thick shell. 
The model of \citet{KM97a} suggests the presence of a molecular cloud 
of $\sim 1\times 10^4\ M_{\odot}$
engulfed by the  blast wave.
The engulfed cloud can act as target material 
for relativistic particles.  
The total (atomic and molecular) hydrogen mass contained 
in the  volume is denoted by 
$M_{\rm H}= \bar{n}_{\rm H} m_p V$. Note that $\bar{n}_{\rm H} \gg n_0$ 
can be expected.

Given the interaction with a molecular cloud, 
we first attribute the observed gamma-rays to
 the decay of $\pi^0$ mesons produced in inelastic collisions between 
accelerated protons (and nuclei) and target gas (Fig.~\ref{fig:SED}).
The gamma-ray spectrum of $\pi^0$-decay 
 is calculated based on \citet{Kamae06} using 
a scaling factor of 1.85 for helium and heavy nuclei \citep{Mori09}. 
Note that the scaling factor assumes the local interstellar abundance for 
target material and the observed cosmic-ray composition. 
Contributions from 
bremsstrahlung and Inverse-Compton (IC) scattering by accelerated electrons 
are also shown in Fig.~\ref{fig:SED}.
 Electron-ion and electron-electron bremsstrahlung spectra are computed 
as in \citet{Baring99}. 
The  interstellar radiation field for IC (see Table~\ref{tbl:model})  is comprised of 
two diluted blackbody components (infrared and optical) and the CMB. 
The infrared and optical components are adjusted to reproduce 
the interstellar radiation field in the GALPROP code \citep{Porter08}.
Cooling effects due to ionization and synchrotron (or IC) losses, 
which introduce cooling breaks in particle spectra in addition to $p_{\rm br}$, 
 are taken into account assuming constant particle injection over 
a period $T_0 \sim {3\times 10^4}$ yr. 
The synchrotron cooling becomes important in the TeV band for leptonic models. 

\begin{figure}[htbp]
  \epsscale{0.8}
    \plotone{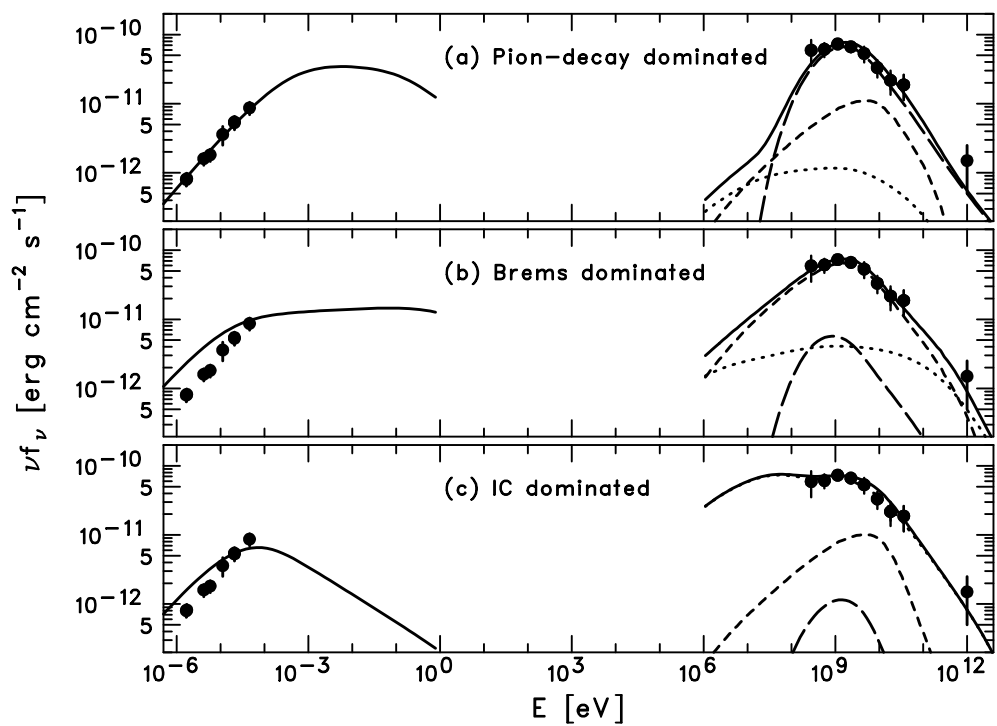}
  \caption{Three different scenarios for the multiwavelength modeling 
  (see Table~\ref{tbl:model}). 
The radio emission \citep[from][]{MK94} is explained by synchrotron radiation, 
while the gamma-ray emission is modeled by different combinations of 
$\pi^0$-decay (long-dashed curve), 
bremsstrahlung (dashed curve), and IC scattering (dotted curve).
The sum of the three component is shown as a solid curve. 
  \label{fig:multiband}}
 \end{figure}

Figure \ref{fig:multiband} (a) shows the radio+gamma-ray spectrum 
together with the radiation model that uses the parameters in Table~\ref{tbl:model}. 
We  adopt here 
 $M_{\rm H} = 2.8\times 10^4\ M_{\odot}$ ($\bar{n}_{\rm H}=10\ {\rm cm^{-3}}$),
which is somewhat larger than the value quoted above. 
The total energy of the high-energy 
protons amounts to $W_p = 5.2\times 10^{50}\ {\rm erg}$, 
which is inversely proportional to $M_{\rm H}$, but insensitive to other parameters. 
The large luminosity of the LAT source can be explained naturally by 
a large number of accelerated protons and dense environments 
as expected for the case of W51C. 

The break feature in the proton spectrum, which is introduced on phenomenological 
grounds,  is not expected in
a shock acceleration zone if acceleration proceeds close to the Bohm limit 
\citep[e.g.,][]{Baring99}. 
Protons can be 
accelerated up to $\sim 200$ TeV  via diffusive shock acceleration 
with a power-law (or even slightly concave) momentum distribution. 
A possible explanation for the falling proton spectrum above 
$p_{\rm br} = 10\mbox{--}30\ {\rm GeV}\, c^{-1}$ (which depends on the choice of 
other parameters)
could be the effects  of damping of 
magnetohydrodynamic turbulence due to ion-neutral collisions. 
According to \citet{PZ03}, 
 the maximum attainable momentum 
in the presence of wave dissipation by ion-neutral collisions is estimated as 
$p_{\rm max}(T_0) \sim 30\, (n/1\, {\rm cm^{-3}})^{-1/2}\ 
{\rm GeV}\, c^{-1}$, using the SNR parameters described above. 
Here $n$ is the ambient neutral gas density. 
The shock precursor cannot confine 
accelerated protons with $p>p_{\rm max}(T_0)$ even if they were accelerated 
earlier. 
Therefore, energy-dependent escape of accelerated particles
from the remnant \citep[e.g.,][]{AA96} may account for the steepening of 
the proton distribution. 
In any case, a break is needed to fit the observed SED. 

In contrast, leptonic models for the GeV gamma-ray production face 
difficulties (see Figure \ref{fig:multiband}b--c and Table~\ref{tbl:model}). 
While the chosen model parameters are not unique in their 
ability to fit the broadband spectrum, they are representative; 
no reasonable choices for them can easily account for the radio+gamma-ray data. 
Leptonic scenarios require $a_e/a_p$ far in excess of cosmic-ray 
abundance ratios. It is difficult to reconcile the 
bremsstrahlung-dominated model 
with the observed radio synchrotron spectrum  (Fig.~\ref{fig:multiband}b).
Also, the IC-dominated model (Fig.~\ref{fig:multiband}c) requires 
an unrealistically large  energy content in
radiating electrons, $W_e \simeq  1\times 10^{51}\ {\rm erg}$, 
a low magnetic field of $B \simeq 2\ \mu{\rm G}$ 
(to  suppress the radio synchrotron emission), and a low density of 
$\bar{n}_{\rm H} < 0.1\ {\rm cm^{-3}}$ (to reduce the bremsstrahlung component). 
Such a low density is problematic because even X-ray-emitting gas has  
a density of $\sim 1\ {\rm cm^{-3}}$ \citep{Koo02}.
It is also difficult to account for the LAT signal by an IC process in 
a possible pulsar wind nebula seen inside W51C,  \CXOJ\ \citep{Koo05}, which has 
only a few parsec extent and 
much weaker radio emission compared with the SNR shell. 

 The most plausible explanation for the LAT source is therefore offered by 
$\pi^0$-decay gamma-rays in a dense 
environment, spawned by efficient acceleration of protons and nuclei 
taking place or having taken place at the shocked shell
of SNR W51C.
The discovery of the GeV-scale gamma-rays presented in this paper 
raises  the intriguing possibility that {\it Fermi} has discerned accelerated ions 
in a middle-aged remnant that is interacting with dense clouds.

\acknowledgments
The {\it Fermi} LAT Collaboration acknowledges support from a number of agencies and institutes for both development and the operation of the LAT as well as scientific data analysis. These include NASA and DOE in the United States, CEA/Irfu and IN2P3/CNRS in France, ASI and INFN in Italy, MEXT, KEK, and JAXA in Japan, and the K.~A.~Wallenberg Foundation, the Swedish Research Council and the National Space Board in Sweden. Additional support from INAF in Italy and CNES in France for science analysis during the operations phase is also gratefully acknowledged.


\clearpage

\begin{deluxetable}{lcccccccc}
\tablecolumns{9}
\tabletypesize{\small}
\tablecaption{Parameters of Multiwavelength Models \label{tbl:model}}
\tablewidth{0pt}
\tablehead{
\colhead{} & \multicolumn{5}{c}{Parameters} & \colhead{}
& \multicolumn{2}{c}{Energetics} \\
\cline{2-6} \cline{8-9} \\
\colhead{Model} & \colhead{$a_e/a_p$} & \colhead{$\Delta s$}
& \colhead{$p_{\rm br}$}  & \colhead{$B$} & \colhead{$\bar{n}_{\rm H}$} & \colhead{} 
& \colhead{$W_p$} 
& \colhead{$W_e$}\\
\colhead{} & \colhead{} & \colhead{} 
& \colhead{(GeV $c^{-1}$)}  & \colhead{($\mu$G)} & \colhead{(cm$^{-3}$)} & \colhead{} 
& \colhead{($10^{50}$ erg)} & \colhead{($10^{50}$ erg)} 
}
\startdata
(a) $\pi^0$-decay 
& 0.02 & 1.4 & 15 & 40 & 10 &
& 5.2 & 0.13 \\
(b) Bremsstrahlung 
& 1.0 & 1.4 & 5 & 15 & 10 &
& 0.54 & 0.87 \\
(c) Inverse Compton 
& 1.0 & 2.3 & 20 & 2 & 0.1 &
& 8.4 & 11 
\enddata
\tablecomments{
Seed photons for IC include
the CMB 
($kT_{\rm CMB} = 2.3\times 10^{-4} $ eV, $U_{\rm CMB} = 0.26\ \rm eV\ cm^{-3}$), 
infrared ($kT_{\rm IR} = 3\times 10^{-3}$ eV, $U_{\rm IR} = 0.90\ \rm eV\ cm^{-3}$), 
and optical  ($kT_{\rm opt} = 0.25$ eV, $U_{\rm opt} = 0.84\ \rm eV\ cm^{-3}$).
The total energy content of radiating particles, $W_{e,p}$, 
is calculated for $p>10\ {\rm MeV}\, c^{-1}$. }
\end{deluxetable}


\end{document}